\documentclass[aps,pra,longbibliography,twocolumn,superscriptaddress,
floatfix]{revtex4-1}
\usepackage[T1]{fontenc}
\usepackage{amsmath}
\usepackage[next]{inputenc}
\usepackage[dvips]{epsfig}
\usepackage{bbm,bm,bbold}
\usepackage{booktabs}
\usepackage{svg} 
\usepackage{multirow}
\usepackage{hhline}
\usepackage{comment}
\usepackage{float}

\usepackage{amsmath,amsfonts,amssymb,amsthm}
\usepackage{color}
\usepackage{graphicx,latexsym}

\usepackage[colorlinks=true,urlcolor=blue,citecolor=blue,linkcolor=blue]{hyperref}

\usepackage{lipsum}

\usepackage{nameref}
\usepackage{varioref}
\usepackage{cleveref}

\usepackage{natbib}

\begin{document}
\renewcommand{\vec}{\mathbf}
\renewcommand{\Re}{\mathop{\mathrm{Re}}\nolimits}
\renewcommand{\Im}{\mathop{\mathrm{Im}}\nolimits}
\newcommand\scalemath[2]{\scalebox{#1}{\mbox{\ensuremath{\displaystyle #2}}}}

\title{Anomalous skew scattering of plasmons in a Dirac electron fluid}

\author{Cooper Finnigan}
\email{cooper.finnigan@monash.edu}
\affiliation{School of Physics and Astronomy, Monash University, Victoria 3800, Australia}

\author{Dmitry K. Efimkin}
\email{dmitry.efimkin@monash.edu}
\affiliation{School of Physics and Astronomy, Monash University, Victoria 3800, Australia}
\affiliation{ARC Centre of Excellence in Future Low-Energy Electronics Technologies, Monash University, Victoria 3800, Australia}

\begin{abstract}
The Berry phase-related nontrivial electronic band geometries can significantly influence bulk and edge plasmons resulting in their non-reciprocal propagation and opening new opportunities for plasmonics. In the present work, we extend the hydrodynamic framework to describe the scattering of plasmons in a Dirac electron fluid off a circular region with an induced nonzero anomalous Hall response, i.e. a Berry flux target. We demonstrate that the scattering has a giant asymmetry or skewness and exhibits a series of resonances. The latter appears due to a chiral non-topological trapped mode circulating the target. We discuss possible experimental realizations, including the surface of a topological insulator film and graphene irradiated by the circularly polarized beam. 
\end{abstract}

\date{\today}
\maketitle

\section{Introdution}

The geometric Berry phase for electrons in solids has been attracting growing interest due to its profound impact on material properties. It is responsible for a variety of phenomena, including polarization, orbital magnetism, various (quantum, anomalous, or spin) Hall effects, photogalvanic effects, and high-harmonic generation (see reviews Ref.~\cite{BFReview1,AHEReview} and references therein). More recently, it has been realized that nontrivial band geometries, characterized by the corresponding electronic Berry curvature,  can significantly influence bulk and edge plasmons, resulting in their non-reciprocal propagation. These phenomena are the most prominent and tunable in materials hosting Weyl and Dirac electronic fluids~\cite{KotovChiralPlasmon,HofmannChiral1,songChiralPlasmonsMagnetic2016,SongChiral2,SongChiral3,VignaleChiralPlasmon} as well as moir\'e superlattices~\cite{NonRecipricalLewandowski1,AgarwalChiral1,NonRecipricalPolini1}. While some of them mimic the effect of the external magnetic field, their remarkable magnitude, unattainable with conventional table-top magnets, opens new avenues for magnetic field-free non-reciprocal plasmonics~\cite{NonRecipricalWeylReview,NonRecipricalMoireReview}.

Recent advances in scattering-type scanning near-field optical microscopy have made it possible to experimentally study the scattering of plasmons in nanostructures~\cite{NearFieldSpectroscopyReview}, drawing significant attention from both experimentalists~\cite{PlasmonScatteringExp1,PlasmonScatteringExp2} and theorists~\cite{PlasmonScattering1,PlasmonScattering2,PlasmonScattering3, PlasmonScattering4,PlasmonScattering5,PlasmonScattering6,PlasmonScattering7,PlasmonScattering8, FinniganEfimkin1,zabolotnykhQuasistationaryNeargatePlasmons2021,valagiannopoulosZeemanGyrotropicScatterers2018,hassanigangarajTopologicalScatteringResonances2020,Valagiannopoulos2008ClosedFormST,2007_Valagiannopoulos}. In a recent study~\cite{FinniganEfimkin1}, we demonstrated that graphene plasmons undergo giant, resonant skew scattering when interacting with a micromagnet. This effect arises from the spatially nonuniform Hall response, prompting an intriguing question: can this phenomenon also occur through the magnetic-field-free anomalous Hall effect, driven by the Berry phase-related nontrivial band geometries of oscillating electrons? 

\begin{figure}[b]
    \centering
    \vspace{-0.2in}
    \includegraphics[width=\columnwidth]{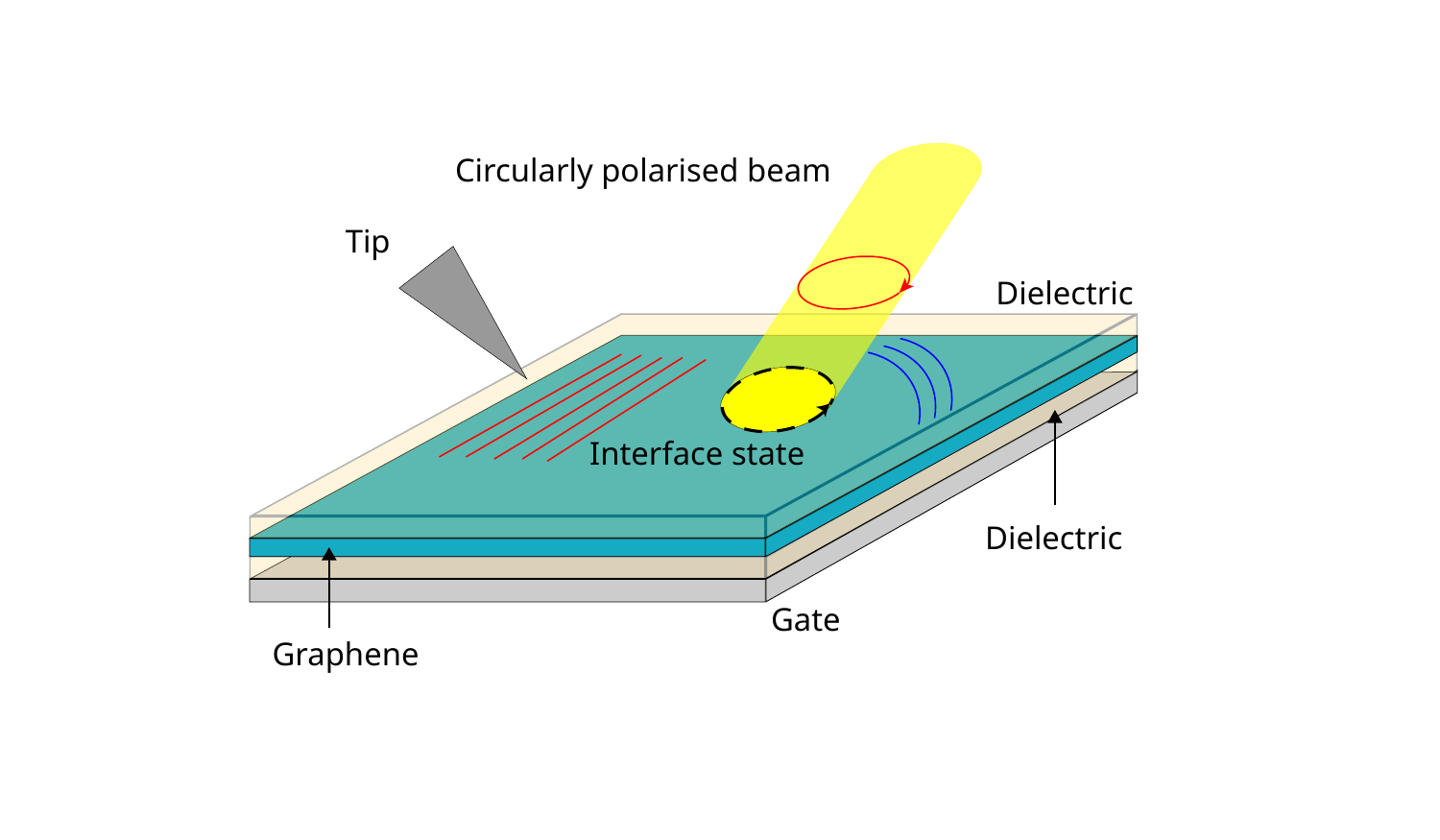}
    \caption{Schematic illustration of the considered setup. Plasmons are excited in graphene by a near-field source and scatter off a non-uniform Berry flux target induced by the circularly polarized high-frequency electromagnetic beam. The scattering exhibits highly skewed and resonant behavior intricately connected to the presence of the chiral plasmon trapped at the Berry flux interface. }
    \vspace{-0.1in}
    \label{Figure1}
\end{figure}

In the present work, we extend the hydrodynamic framework~\cite{songChiralPlasmonsMagnetic2016} to describe the scattering of plasmons in a Dirac electron fluid. For a target, we consider a circular region with an induced nonzero anomalous Hall response, i.e. a Berry flux target. As sketched in Fig.~\ref{Figure1}, the setup can be engineered in graphene pumped by the circularly polarized high-frequency electromagnetic beam~\cite{cayssolFloquetTopologicalInsulators2013,okaPhotovoltaicHallEffect2009,KitagawaFloquet,GrapheneFloquetExp1}. Furthermore, a non-zero anomalous Hall response can also be engineered at the surface of a topological insulator with deposited ferromagnetic flake ~\cite{ProximityInducedFerromagnetismTI,ProximityInducedFerromagnetismTI2,FuhrerReview}. We demonstrate that plasmons experience giant and resonant skew scattering due to a trapped chiral plasma mode circulating the flake/beam. Although certain aspects of this behavior resemble the scattering problem involving a micromagnet target, the anomalous Hall response favors high resonance frequencies that conventional micromagnets cannot achieve.\\
The rest of the paper is organized as follows. Sec.~\ref{HydroFramework} presents the hydrodynamic framework to describe plasmons in the regime of anomalous Hall effect. Sec.~\ref{BerryFluxInterface} is devoted to deriving the chiral interface mode at the Berry flux interface. Sec.~\ref{ScatteringTheory} presents the necessary scattering theory and results. Finally, Sec.~\ref{Discussion} presents estimations relevant to material realizations and broad discussions therein.
\section{Hydrodynamic framework}
\label{HydroFramework}
The long-wavelength dynamics of plasmons in a Dirac electron fluid can be described by modified collisionless hydrodynamic equations~\cite{songChiralPlasmonsMagnetic2016}, which properly account for nontrivial band geometry of the Dirac electrons via an anomalous velocity contribution. They can be obtained via coarse-graining of semiclassical equations of motion for an individual electron
\begin{equation}
\mathbf{v}_{\mathbf{p}}=\frac{\partial \varepsilon_{\mathbf{p}}}{\partial\vec{p}}
+\frac{1}{\hbar} \left[\dot{\mathbf{p}} \times \boldsymbol{\Omega}_{\mathbf{p}}\right], \quad \quad \dot{\mathbf{p}}=e \nabla \phi(\mathbf{r}). \label{Eq.0}
\end{equation}
Here $\phi(\mathbf{r})$ is the electric potential, and $e>0$ is the absolute value of the electron charge and $\varepsilon_{\mathbf{p}}$ is the electronic band dispersion. The Berry curvature $\boldsymbol{\Omega}_{\mathbf{p}}=\Omega_{\mathbf{p}}\mathbf{e}_{z}$, the local measure of the band geometry, induces the additional anomalous velocity and ensures frequency-independent and magnetic field-free Hall response of the electron fluid. The resulting coarse-grained equations for  electron charge density, $\rho(\mathbf{r},t)$, and current density, $\mathbf{j}(\mathbf{r},t)$, are given by
\begin{equation}
    \begin{gathered}
    \partial_t \rho(\mathbf{r}, t)+\nabla \cdot \mathbf{j}(\mathbf{r}, t)=0, \\
    \partial_{t}\mathbf{j}(\mathbf{r},t)=-\frac{n e^{2}}{m}\nabla \phi(\mathbf{r},t)- \frac{e\mathcal{F}}{\hbar}\left[ \mathbf{e}_{z} \times \nabla \dot{\phi}(\mathbf{r},t) \right]. \label{Eq.1}
    \end{gathered}
\end{equation}
Here, $m$ is the collective (or cyclotron) mass of electrons, and $n$ is their equilibrium concentration; $\mathcal{F}$ is the total Berry flux in
momentum space, which pierces all occupied electronic states. The Eqs.~(\ref{Eq.1}) need to be supplemented with the Poison equation, which self-consistently connects the scalar potential with the oscillating charge density. In the presence of a metallic gate, the scalar potential created by the charge-density fluctuations
becomes overscreened and can be described using the local capacitance approximation, i.e. $\phi(\mathbf{r},t)= \rho(\mathbf{r},t)/C$. Here  $C=\kappa/2 \pi d$ is the capacitance per unit area, $d$ is the distance to the gate, and $\kappa$ is the dielectric constant of the surrounding medium.

The hydrodynamic equations support plasmons with a linear dispersion $\Omega(\mathbf{q})=u q$ and a plasmon velocity of $u=\sqrt{2 \pi n e^2 d / m \kappa}$. It has already been noticed that the bulk dynamics of plasmons are not affected by the anomalous Hall response of the Dirac fluid~\cite{songChiralPlasmonsMagnetic2016}. To see this, Eqs.~(\ref{Eq.1}) can be combined into single wave equation
\begin{equation}
\partial_t^2 \rho(\mathbf{r}, t)-u^2 \nabla^2 \rho(\mathbf{r}, t)=0, \label{WaveEquation}
\end{equation}
which is independent of the Berry flux. The latter introduces a transverse component of electric current oscillations and, therefore, strongly alters the physics of edge plasmons at the sample boundaries~\cite{songChiralPlasmonsMagnetic2016,kumarChiralPlasmonGapped2016,smithEdgeModesFabryPerot2019}. 

The modified hydrodynamic equations can describe the scattering setup illustrated in Fig.~(\ref{Figure1}). The role of the target is played by a disk-shaped region with finite net Berry flux, $\mathcal{F}(r) = \mathcal{F}\Theta(r_\mathrm{F}-r)$, where $\mathcal{F}$ is the magnitude of the Berry flux induced by the circularly polarized beam and $r_\mathrm{F}$ is its radius. The flux induction necessitates the opening of a gap in the Dirac spectrum, leading to electron depletion and a reduction in the plasmon velocity within the target. However, for realistic parameters, the velocity mismatch has a negligible effect on the resonant skew scattering of plasmons. Although the main text neglects this mismatch for the sake of simplicity, it is revisited and analyzed in detail in Appendix~\ref{MaterialRealizations}. Before delving into the scattering problem, it is useful to first examine the Berry flux interface, a key feature of the target.

\section{Chiral interface mode}
\label{BerryFluxInterface}

To get insights into the physics of plasmons at the Berry flux interface, it is instructive to consider first a half-plane geometry before the disk geometry.  As demonstrated in Appendix~\ref{InterfaceStates}, the interface, i.e. $\mathcal{F}(x) = \mathcal{F}\Theta(x)$, traps a single chiral mode with the dispersion relation 
\begin{equation}
    \Omega_{\mathrm{int}}(q_{y})= \frac{u q_{y} \omega_{*}}{\sqrt{(u q_{y})^{2}+\omega_{*}^{2}}}. \label{Eq.3}
\end{equation}
This mode is chiral and non-topological, coexisting with the bulk modes. 
The low frequency behaviour is linear, $\Omega_{\mathrm{int}}(q_{y}\ll \omega_*/u)\approx u q_y$, and then saturates to $\Omega_{\mathrm{int}}(q_{y}\gg \omega_*/u)\approx\omega_*$. The physical meaning of the threshold frequency $\omega_{*} = 2n \hbar / m \mathcal{F}$ can be elucidated by comparing longitudinal (normal) and transverse (anomalous) contributions to the electric current. For frequencies below $\omega_{*}/2$, the longitudinal response dominates, whereas the transverse response becomes dominant at higher frequencies.

\begin{figure}[t]
    \vspace{0.5cm}
    \centering
    \includegraphics[width=1\columnwidth]{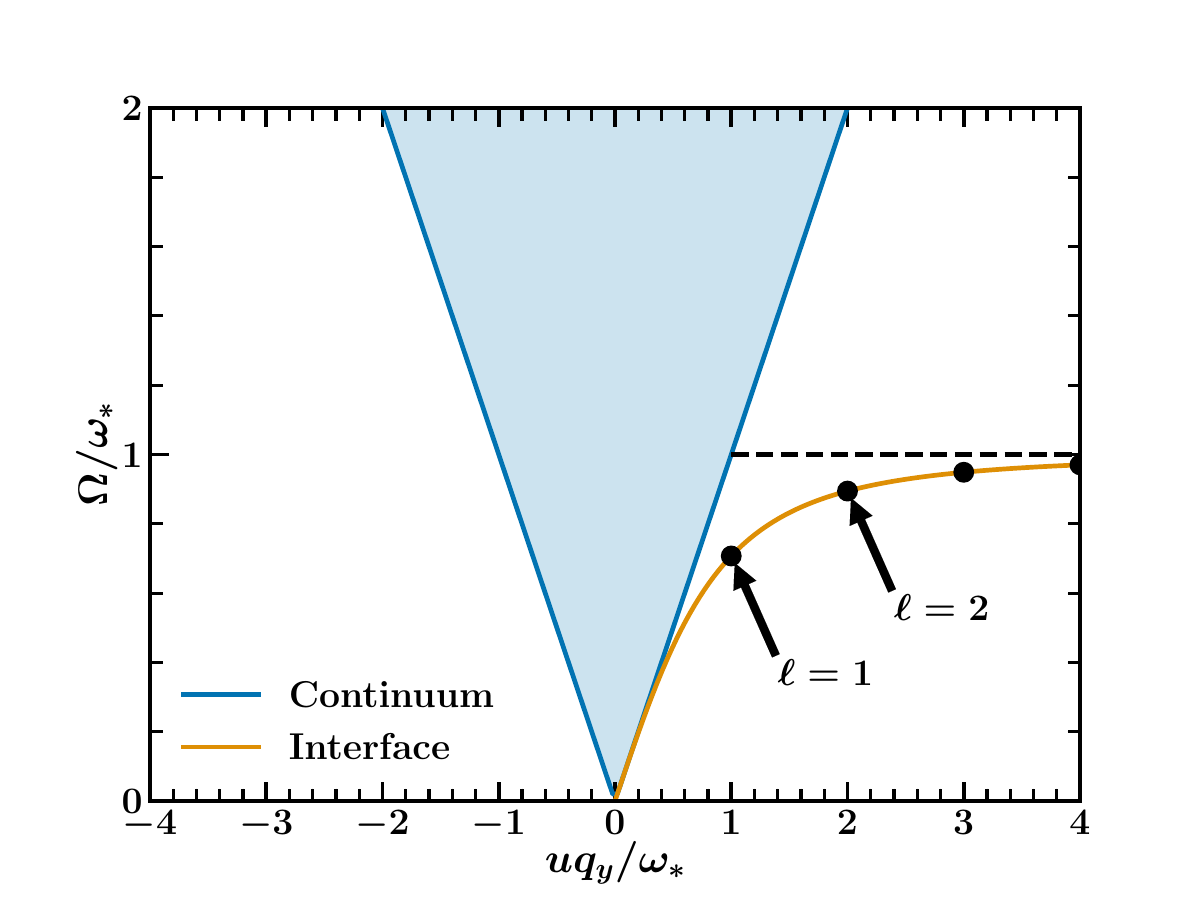}
    \caption{The dispersion of the interfacial plasmon (orange), trapped at the Berry flux interface, is described by Eq.~(\ref{Eq.3}). This interface mode lies below the continuum of bulk states (blue) and exhibits unidirectional wave propagation, making it chiral. Resonant frequencies for a circular interface calculated for $\alpha = 1$ are given by Eq.~(\ref{Resonantfrequncies}) and are marked as black dots. At high wavenumbers, the dispersion saturates towards the frequency $\omega_{*}$, indicated by the black dashed line. }
    \label{Figure 2}
\end{figure}

The penetration length of the interface states is wave-vector-dependent and is given by 
\begin{equation}
l_{\mathrm{int}}(q_{y}) = \frac{\sqrt{(u q_{y})^{2}+\omega_{*}^{2}}}{u q_{y}^{2}}. 
\end{equation}
The penetration length rapidly grows with $q_y$, reflecting that the stability of the interface modes depends on how well they separate from the continuum of bulk modes. As seen in Fig.~(\ref{Figure 2}) the interface mode is well separated and anticipated to be stable in the frequency range between $\omega_*/2$ and $\omega_*$.      

In the circular geometry sketched in the setup Fig.~(\ref{Figure1}), the interface modes are subject to the standing wave condition $2\pi q_{\ell} r_{\mathrm{F}}=2\pi \ell$, where $\ell=1, 2, 3,...$ is an integer. Therefore, the allowed interface frequencies become discrete as 
\begin{equation}\Omega_{\ell}=\frac{\ell \omega_{\mathrm{g}}\omega_{*}}{\sqrt{(\ell \omega_{\mathrm{g}})^2 + \omega_{*}^2}},
\label{Resonantfrequncies}\end{equation} where $\omega_{\mathrm{g}}=u/r_\mathrm{F}$ determines the splitting between the trapped modes with the lowest frequencies. These discrete chiral states reveal themselves via resonant skew scattering events, which is the core result of this paper.  
\section{Scattering theory and results} \label{ScatteringTheory}
In the considered setup, disk-shaped Berry flux plays the role of the target. Due to the azimuthal symmetry of the target, the scattering problem can be addressed using partial wave analysis. The asymptotic behavior for the electron density satisfying Eq.~(\ref{WaveEquation}) can be written as  
\begin{equation}
\rho(r, \phi) =e^{i k r \cos \phi}+f(\phi)\frac{e^{i k r}}{\sqrt{ i r}}. \label{Partial_Waves}
\end{equation}
The first term describes an incident and passed plane wave,
whereas the second term represents the scattered wave. The scattering amplitude $f(\phi)$ can be presented as  
\begin{equation}
f(\phi)=\sqrt{\frac{1}{2 \pi q}} \sum_\ell e^{i \ell \phi}\left[1-e^{-2 i \delta_\ell(q)}\right], \label{eq 3.13}
\end{equation}
\begin{figure}[t]
    \centering
    \includegraphics[clip,trim={0cm 0cm 0cm 0cm}, width=\columnwidth]{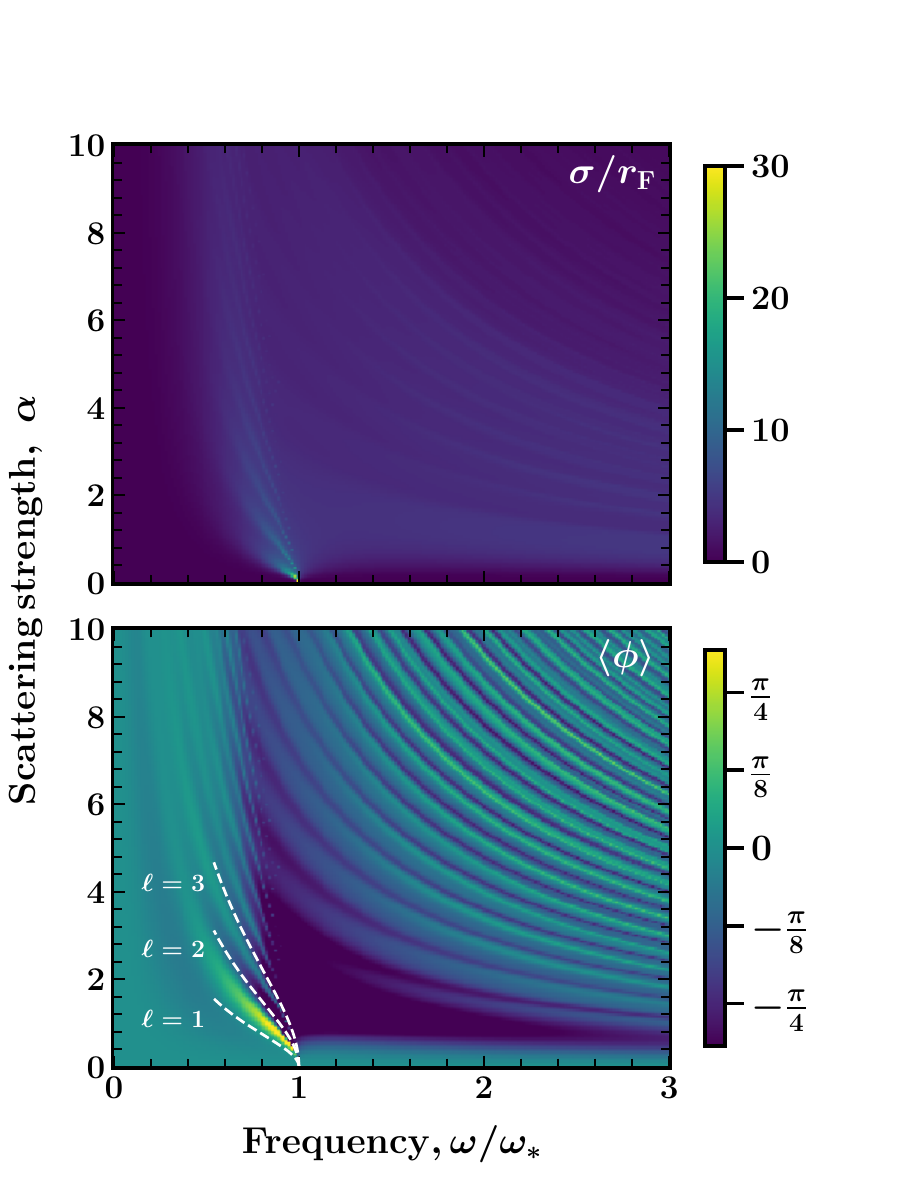}
    \caption{The scattering strength and frequency dependence of the total cross-section $\sigma$ (top) and the average scattering angle $\langle \phi \rangle$ (bottom) are presented. The bottom subplot additionally highlights the first three resonant harmonics $(\ell = 1, 2, 3)$, determined by Eq.~(\ref{Resonantfrequncies}) and shown as white dashed lines. }
    \label{Figure 3}
\end{figure}
where $\delta_{\ell}(q)$ is the phase shift for the partial wave labeled by the discrete orbital number $\ell$. These phase shifts can be calculated using the radial equation for $\rho(r,\phi)$ supplemented with a pair of boundary conditions. The first one dictates the continuity of electric potential and is not sensitive to the presence of the Berry flux jump. 
\begin{figure}[h]
    \centering
    \includegraphics[clip,trim={0cm 0cm 0cm 0cm}, width=\columnwidth]{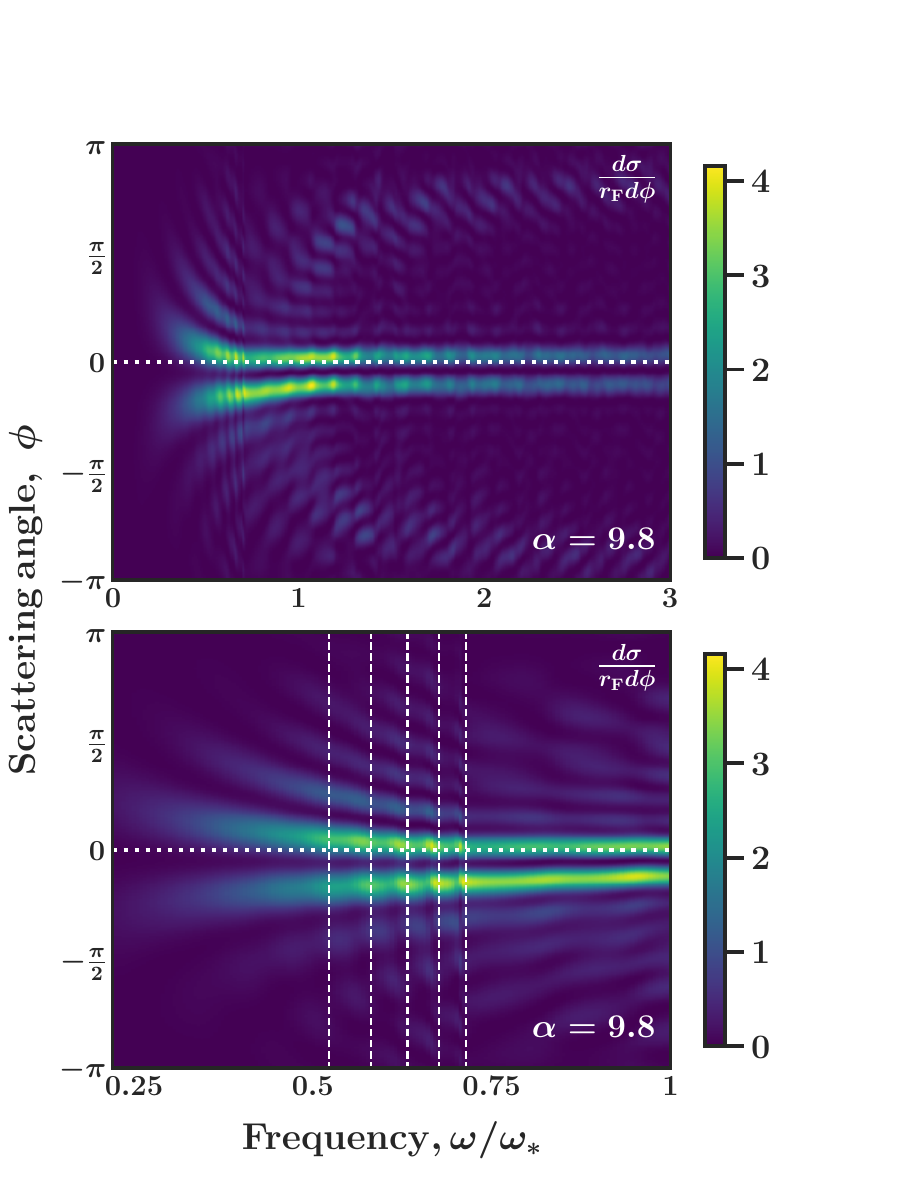}
\caption{The scattering angle and frequency dependence of the differential cross section $d\sigma/d\phi$, calculated for scattering strength $\alpha=9.8$, corresponding to the graphene-based setup. The bottom panel zooms at the frequency range with multiple scattering resonances, which are given by Eq.~(\ref{Resonantfrequncies}) and are highlighted as vertical dashed lines.}
    \label{Fig_resubmission}
\end{figure}
The second one ensures radial current density continuity at the interface and incorporates the Berry flux-induced anomalous contribution to electric current.  The Berry flux enters via the dimensionless combination $\alpha = r_{\mathrm{F}}\omega_{*}/u$, which can be interpreted as the dimensionless scattering strength. These calculations are presented in Appendix~\ref{PartialWaves}, whereas this main text of the paper focused on an analysis of the scattering observables.

The angular distribution of the plasmon scattering is determined by the differential cross section $d\sigma/d\phi = |f(\phi)|^{2}$. The scattering probability and skewness can be characterized by the total cross-section $\sigma$ and average scattering angle $\langle\phi\rangle$, which are given by 
\begin{equation}
\sigma=\int_{-\pi}^\pi|f(\phi)|^2 d \phi, \quad \langle\phi\rangle=\frac{1}{\sigma} \int_{-\pi}^\pi \phi|f(\phi)|^2 d \phi. \label{Eq.6}
\end{equation}
Their dependencies on the dimensionless frequency of the incoming plasmon $\omega/\omega_*$ and the controlling scattering parameter $\alpha$ are presented in Fig.~(\ref{Figure 3}).

At the relatively high frequencies $\omega\gtrsim \omega_*$, the total cross-section $\sigma$ experiences slightly visible oscillations, common in scattering problems with a sharp profile of the target and approximately follow the squared Fourier image of the target. As for the average scattering angle $\langle\phi\rangle$, its oscillations are more prominent and involve a sign change. At relatively low frequencies, $\omega \lesssim \omega_*/2$, the total cross-section $\sigma$ and the average scattering angle $\langle\phi\rangle$ become negligibly small and exhibit no distinct features. This behavior reflects the fact that the anomalous term in Eq.~(\ref{Eq.1}) is proportional to the time derivative of the scalar potential and diminishes as the frequency decreases.  \\
In the intermediate frequency regime $\omega_*/2\lesssim \omega \lesssim \omega_*$, the frequency dependence of the total cross-section $\sigma$ exhibits a few narrow peaks. As clearly seen in Fig.~(\ref{Figure 3}), their frequencies are accurately described by the expression for Eq.~(\ref{Resonantfrequncies}), confirming that the resonant behavior is intricately connected with the chiral plasmon trapped at the Berry flux interface and circulating the target. The average scattering angle $\langle\phi\rangle$ achieves a maximum below the resonance, rapidly switches its sign, and then has a prominent minimum. Its extrema are giant and exceed $\pm \pi/4$. The giant resonant scattering induced by the Berry flux target is the main result of this paper.

The resonant features described above can also be observed in the angular distribution and frequency dependence of the differential scattering cross-section, $d\sigma/d\phi$. Fig.~(\ref{Fig_resubmission}) presents these dependencies for a scattering strength of $\alpha = 9.8$, corresponding to the graphene-based setup (see Sec.~\ref{Discussion} for estimations).  At low frequencies, $\omega \lesssim \omega_{*}$, the resonant scattering frequencies given by Eq.~(\ref{Resonantfrequncies}) manifest as alternating maxima and minima in the angular distribution, as indicated by the dashed vertical lines in the bottom panel. At high frequencies, $\omega \gtrsim \omega_{*}$, forward scattering becomes dominant, as expected. However, an asymmetry persists around the purely forward direction, as shown in the top panel.

\section{Estimations and discussions} \label{Discussion}
The anomalous Hall response in both considered setups can be described by the massive Dirac model. The corresponding Hamiltonian is given by $H=v \sigma \cdot \hat{\vec{p}}+\Delta \sigma_z$, where $\sigma$ is the electron (pseudo-)spin operator~\footnote{For Dirac electrons at the surface of topological insulator, the spin operator vector needs to be rotated by $\pi/2$} and $v$ is the Dirac velocity. The Dirac mass $\Delta>0$ is assumed to be nonzero within the circular target region. The corresponding Berry flux is $\mathcal{F}=g \Delta/4\pi \epsilon_\mathrm{F}$ and collective mass is $m=\epsilon_\mathrm{F}/v^2$ have a strong doping dependence parametrized by electronic Fermi energy $\epsilon_\mathrm{F}$ and the degeneracy factor $g$ for the Dirac electrons. 

The resulting characteristic frequency is given by 
$\hbar\omega_* = 2 (\epsilon_\mathrm{F}^2 - \Delta^2)/\Delta
$~\footnote{Interestingly, this expression is more generic and valid for chiral charge carriers in graphene multilayers. See Appendix~\ref{GrapheneMultilayers} for discussion}. It is determined exclusively by the electronic energy scales that favor the high-frequency nature of the scattering resonances. This naturally raises the question of whether the hydrodynamic theory we employ above can describe the resonant plasmon scattering. As we discuss in Appendix~ \ref{HydroKubo}, the hydrodynamic theory accurately describes both longitudinal and transverse conductivities in the wide frequency range $\hbar\omega\lesssim\epsilon_{\mathrm{F}}$~\footnote{It should be mentioned that predictions of the hydrodynamic theory reasonably agree with the microscopic one up to even higher frequencies, approximately $3\epsilon_\mathrm{F}/2$}. All resonances are within this range if $\Delta \gtrsim 0.78 \epsilon_\mathrm{F}$, while only the lowest ones are accurately captured for the least strict condition $\Delta \gtrsim 0.62 \epsilon_\mathrm{F}$. These conditions provide expectations for the magnitude of the Dirac mass, which is responsible for the local AHE. They can be satisfied for both considered setups. 

The first setup involves graphene irradiated with circularly polarized light. If the frequency is high enough and non-resonant, electron interband transitions and heating effects are minimized. Instead, the Dirac mass is coherently generated~\cite{cayssolFloquetTopologicalInsulators2013,okaPhotovoltaicHallEffect2009,KitagawaFloquet,GrapheneFloquetExp1} and can be approximated to be constant inside the beam. For estimations, we use $\Delta\approx30~\hbox{meV}$ and $\epsilon_\mathrm{F}\approx 45\;\hbox{meV}$. The resonances are in the range between $38\sim 76\;\hbox{meV}$ and their penetration lengths are $10~$$\sim30~\hbox{nm}$. For a beam size of $r_\mathrm{F}=100\; \hbox{nm}$, the characteristic splitting between resonances and the dimensionless scattering strength are given as $\hbar\omega_\mathrm{g}\approx 7.8\;\hbox{meV}$ and $\alpha\approx 9.8$ respectively. In this frequency range plasmons in graphene are typically  observed~\cite{GraphenePlasmonicsReview1}. 

The second setup involves a topological insulator thin film placed at a massive gate. The local anomalous Hall response at its top surface~\footnote{The Dirac electrons at the bottom surface of the film are assumed to hybridize with conducting electrons in the gate} can be induced via the magnetic proximity effect with a ferromagnetic flake, e.g. $\hbox{Tm}_3\hbox{Fe}_5\hbox{O}_{12}$ or $\hbox{Y}_3 \hbox{F}_5 O_{12}$ \cite{ProximityInducedFerromagnetismTI,ProximityInducedFerromagnetismTI2,FuhrerReview}. The proximity-induced Dirac mass up to $\Delta\approx50~\hbox{meV}$ has been reported. For Fermi energy $\epsilon_\mathrm{F}\approx 70\;\hbox{meV}$, resonances are in the range $48$ $\sim 96\;\hbox{meV}$ and their penetration length is $5~$$\sim20~\hbox{nm}$. For the flake radius $r_\mathrm{F}=20\; \hbox{nm}$, the characteristic splitting between resonances and the dimensionless scattering strength are $\hbar\omega_\mathrm{g}\approx 23\;\hbox{meV}$ and $\alpha\approx 4.2$. This is also the frequency in which plasmons at the surface of topological insulators have been previously reported~\cite{TIPlasmonExp1,TIPlasmonExp2, ginleyDiracPlasmonsPresent2018,stauberPlasmonicsTopologicalInsulators2017}

The predicted giant resonant skew scattering shares notable similarities with the scattering problem involving a micromagnet. Specifically, both magnetic field and Berry flux interfaces trap chiral plasmons, and the functional forms of their corresponding dispersion relations are identical~\footnote{The dependence of the threshold frequency on the strength of the transverse response exhibits opposite trends for the two problems, resulting in distinct behaviors of the plasmon spectra as the interface emerges. For the magnetic field interface, $\omega_\mathrm{c} \sim B$, and the trapped chiral mode arises from the zero-frequency band, favoring the low-frequency nature of the scattering resonances. Conversely, for the Berry flux interface, $\omega_* \sim \mathcal{F}^{-1}$, and the trapped chiral mode originates from the positively dispersive band, favoring the high-frequency nature of the resonances. These contrasting behaviors reflect the frequency dependence of the conductivity tensor. In the conventional Hall effect, the transverse response dominates at low frequencies, $\omega \lesssim \omega_\mathrm{c}$. In contrast, for the anomalous Hall effect, the transverse response becomes dominant at higher frequencies, $\omega \gtrsim \omega_*$}. In this analogy, $\omega_*$ is replaced by $\omega_\mathrm{c}/2$, where $\omega_\mathrm{c} = eB / mc$ represents the cyclotron frequency associated with the magnetic field $B$ generated by the micromagnet. The scales, however, are drastically different: the resonances for the micromagnet problem appear in the sub-meV range, and the corresponding penetration length of chiral trapped states is at the $\mu\hbox{m}$ scales. The two orders of magnitude difference compared to the scales for the Berry flux target cannot be breached using conventional table-top micromagnets.

For the problem of plasmon scattering off a micromagnet, the hydrodynamic equations can be reformulated as an effective spin-1 Dirac problem with a spatially dependent Dirac mass~\cite{TopologicalMagnetoplasmon2016,FinniganKargarianEfimkin1,FinniganEfimkin1}. This reformulation leverages scattering theories originally developed for electronic systems and establishes connections with electronic scattering phenomena in nanostructures~\cite{SkewScattering1,SkewScattering2,TISkewSkyrmion1,TISkewSkyrmion2,KleinScattering1,KleinScattering2}. As outlined in Appendix~\ref{Heff}, the chiral hydrodynamic equations also permit an eigenvalue reformulation. However, the associated effective Hamiltonian is non-Hermitian~\footnote{The effective Hamiltonian respects the interplay between inversion and particle-hole symmetries, ensuring the stability of its eigenvalues, which describe the dispersion relations of plasmons.}. This alternative representation does not appear to provide any immediate computational or interpretative advantages for scattering results. 

Skew scattering of plasmons can also be approached microscopically \cite{HerbFertigPaper}. In particular, quantum plasmon-carrying states in the regime of the AHE were argued to have a quantum geometric dipole and exhibit plasmon skew scattering off a density inhomogeneity. In contrast to the problem considered here, the scattering target does not break time-reversal symmetry. This problem can also be approached by the hydrodynamic framework developed here but is left to future studies.

The scattering problem under consideration has an inverse counterpart \cite{Li:23}, which focuses on reconstructing the spatial profile of the Berry flux based on the angular and frequency dependence of plasmon scattering. The resonant skews scattering, however, relies on the presence of chiral interface plasma mode and is not sensitive to the shape or spatial profile of the Berry flux target \footnote{See the supplementary material of Ref.~\cite{FinniganEfimkin1} for calculations showing the robustness of skew scattering to magnetic scatterer shape, and the robustness of the interface mode to the profile of the magnetic field profile variations.}. Therefore, the inverse aspect of the problem lies beyond the scope of the present paper. 

Within the employed local capacitance approximation, the hydrodynamic description of plasmons can be mapped to shallow water hydrodynamics. The mathematical structure of the latter makes it well-suited for applying transformation optics techniques and concepts from hydrodynamic metamaterials \cite{HydroTransOptics,HydroMeta1,HydroMeta2,HydroMet3}. Extending these approaches to plasmon dynamics in systems with spatially nonuniform Berry flux for oscillating charge carriers would provide an alternative to traditional transformation optics methods \cite{GrapheneTransOptics}, which are primarily based on Maxwell's equations and designed for controlling plasmons in two-dimensional materials and metallic surfaces.

We have focused on plasmon scattering by a single, isolated Berry flux target. However, gratings or patterns composed of multiple Berry flux targets could potentially be utilized to control and enhance plasmon propagation. These structures would complement recently developed approaches, such as spatial modulation of electron density and dielectric properties in the surrounding graphene environment \cite{SPP_1,SPP_2,SPP_3}, as well as the formation of vertical and multilayer structures incorporating other monolayer materials \cite{lowPolaritonsLayeredTwodimensional2017}.

In conclusion, we extended the hydrodynamic framework to describe the scattering of plasmons in Dirac electronic systems with nontrivial band geometries. When a nonzero Berry flux acts as the scattering target, plasmons exhibit giant, resonant scattering. We discuss potential material implementations, such as the surface of a topological insulator with a deposited ferromagnetic flake and graphene driven by a circularly polarized electromagnetic beam. Furthermore, our approach can be generalized to other scattering scenarios involving plasmons. 
\section*{Acknowledgements}
We acknowledge fruitful discussions with Michael Fuhrer, Haoran Ren, Agustin Schiffrin, and Gary Beane as well as support from the Australian Research Council Centre of Excellence in Future Low-Energy Electronics Technologies (CE170100039). 
\appendix
\section{Material realizations}
\label{MaterialRealizations}
\subsection{Graphene pumped by circularly polarized beam}
As for the first setup, Graphene is placed near a gate, and a small circular region is pumped with circularly polarised light (CPL). The CPL induces a local anomalous Hall response and an effective Dirac mass in this region. We estimate the following set of parameters for this setup: electron velocity outside of the irradiated region $v \approx 1 \times 10^{8} \ \mathrm{cm}/\mathrm{s}$, electron concentration outside of the irradiated region $n \approx 1.5 \times 10^{11} \ \mathrm{cm}^{-2}$, distance to the gate $d \approx 18 \mathrm{nm}$, dielectric constant for the surrounding dielectric medium ($\mathrm{SiO}_{2}$) of $\kappa \approx 4$, and a CPL induced Dirac mass of $\Delta =30 \ \mathrm{meV}$. The plasmon velocity inside and outside the Berry flux target are $u_{\mathrm{out}}=1.2 \times 10^{8} \ \mathrm{cm}/\mathrm{s}$ and $u_{\mathrm{in}} = 8.6 \times 10^{7} \ \mathrm{cm}/\mathrm{s}$. The relevant frequency scales for the scattering theory calculations are $\hbar \omega_{*} =75 \ \mathrm{meV}$, $\hbar \omega_{\mathrm{th}}=90 \ \mathrm{meV}$, and $\hbar \omega_{\mathrm{g}}=7.8 \ \mathrm{meV}$ for a spot radius of $r_{\mathrm{F}}=100 \ \mathrm{nm}$. 
\subsection{The surface of topological insulator}
As for the second setup, a thin topological insulator film is placed at the massive gate. The local anomalous Hall response at its top surface~\footnote{The Dirac electrons at the bottom surface of the film are assumed to hybridize with conducting electrons in the gate} can be induced via the magnetic proximity effect with magnetic flake with out-of-surface magnetization. For estimations, we use the following set of parameters: electron velocity $v\approx 6.6\times 10^{7} \;\hbox{cm}/\hbox{s}$, electron concentration outside the flake $n\approx 2.0 \times 10^{11}\; \hbox{cm}^{-2}$, distance to the gate $d\approx50\;\hbox{nm}$, effective optical dielectric constant of the surrounding medium  $\kappa\approx 12$, and the magnetic proximity induced Dirac mass $ \Delta=50\;\hbox{meV}$. The plasmon velocity inside and outside the flake $u_\mathrm{out}=6.9\times 10^{7}\; \hbox{cm}/\hbox{s}$ and $u_\mathrm{in}=4.8 \times 10^{7}\; \hbox{cm}/\hbox{s}$. The difference frequencies in the scattering theory are given by $\hbar\omega_*=96 \; \hbox{meV}$, $\hbar\omega_{\mathrm{th}}=140\;\hbox{meV}$ and $\hbar\omega_\mathrm{g}=23~\hbox{meV}$ for a flake radius $r_\mathrm{F}=20 \hbox{nm}$. The system hosts a few well-separated resonances with the penetration length below $5 \hbox{nm}$.
\subsection{Scattering results}
\begin{center}
    \begin{figure}[h]
     \includegraphics[trim = {0cm 0.5cm 1cm 1.8cm}, width=0.95\columnwidth]{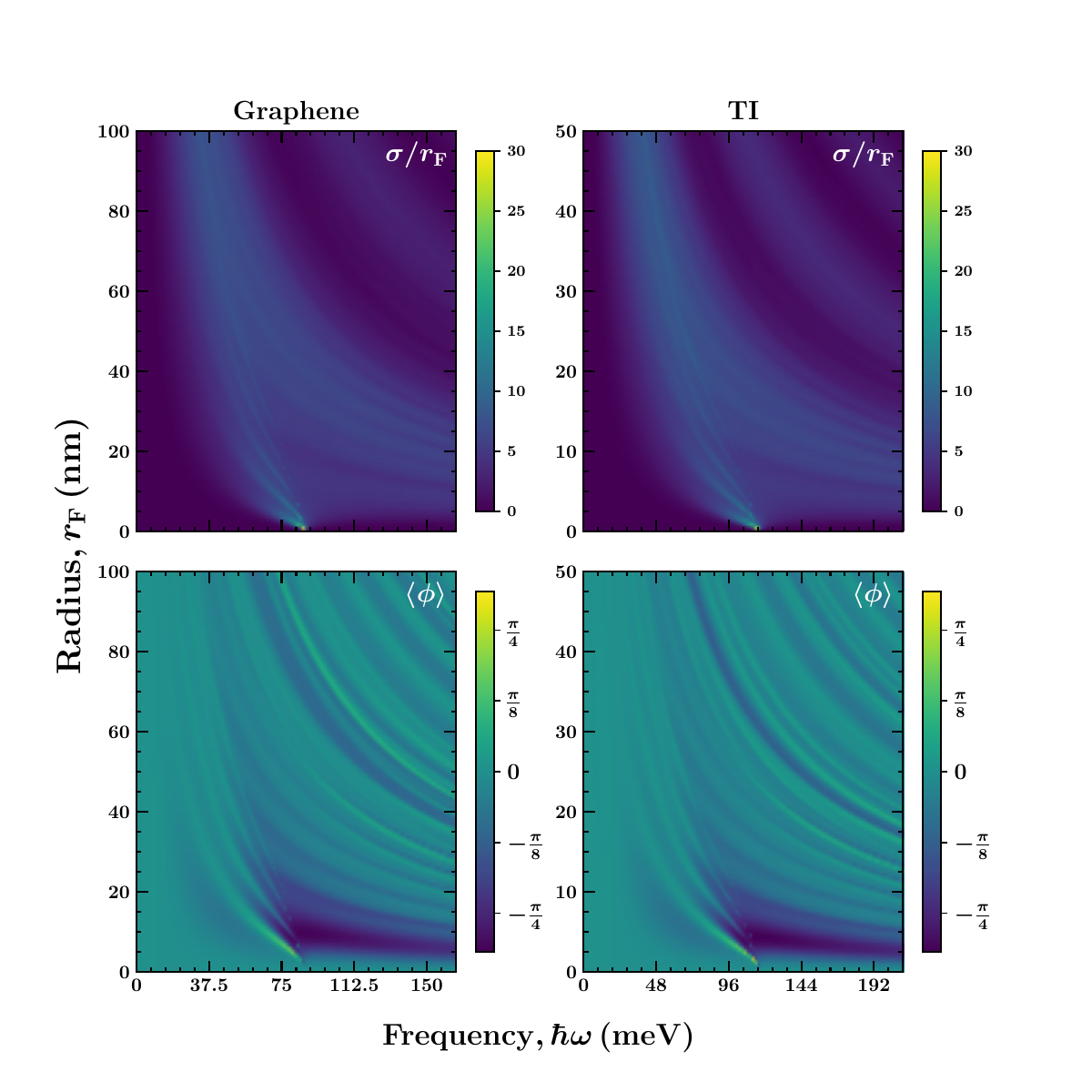}
    \caption{Beam/flake radius and frequency dependence of the total cross sections $\sigma$ (top row) and the average scattering angle $\langle \phi \rangle$ (bottom row) for the Dirac electrons in graphene (left column) and at the surface of topological insulator (right column). The calculations account for the plasmon velocity mismatch, which was neglected in the main part of the paper and has a minimal impact on the resonant skew scattering caused by the chiral plasmon trapped at the Berry flux interface. Instead, it primarily influences the high-frequency oscillations in a manner similar to the scattering problem involving a micromagnet~\cite{FinniganEfimkin1}.}
    \label{Figure_5}
    \end{figure}
\end{center}

Beam/flake radius and frequency dependence of the total cross sections $\sigma$ and the average scattering angle $\langle \phi \rangle$ for the Dirac electrons in graphene and at the surface of topological insulator are presented in Fig.~(\ref{Figure_5}). The calculations account for the plasmon velocity mismatch, which was neglected in the main part of the paper. The mismatch has minimal impact on the resonant skew scattering caused by the chiral plasmon trapped at the Berry flux interface. Instead, the mismatch impacts the high-frequency behavior of scattering observables in a manner similar to the scattering problem involving a micromagnet~\cite{FinniganEfimkin1}. 
\section{The chiral interface plasmon}
\label{InterfaceStates}
This Appendix presents the derivation of the interface mode at the Berry flux interface, i.e. $\mathcal{F}(x)=\mathcal{F}\Theta(x)$. If we take into account the translational invariance in the y-direction, the localized solution of the wave equation for electron density, Eq.~(\ref{WaveEquation}), can be presented as

\begin{equation}
\rho(x<0)=\rho_0^- e^{i q y + \kappa x}, \quad \; \rho(x>0)=\rho_0^+ e^{i q y -\kappa x}. 
\end{equation}

Here $\rho_0^\pm$ are constants, and $\kappa$ is the inverse decay length of the interface mode given by 
\begin{equation}
\kappa=\frac{\sqrt{u^2 q^2-\Omega_q^2}}{u}. \label{A.2}
\end{equation}
It depends on the wave vector along the interface $q$ and a dispersion relation of the mode $\Omega_q$. The problem needs to be supplemented with a pair of boundary conditions. The first one dictates the continuity of electric density (it guarantees the continuity of electric potential) and is satisfied if $\rho_0^-=\rho_0^+$. It is not sensitive to the presence of the Berry flux jump. 
The second one ensures electron charge conservation at the interface and is satisfied if the x-component of electric current  $j_x$ is continuous at the interface. It includes both normal and anomalous components 
\begin{equation}
j_x=-i \frac{u^2}{\Omega_q} \partial_x \rho+\frac{2 u^2}{\omega_*} \Theta(x) \partial_y \rho.
\label{TransverseCurrent}
\end{equation}
and is continuous at the interface if $\kappa \omega_*=u q \Omega_q$. For $\mathcal{F}>0$ ($\mathcal{F}<0$), the equation has a solution only for $q>0$ ($q<0$) that provides the interface mode with its chiral nature. Its explicit solution results in the dispersion relation presented in the main part of the paper as Eq.~(\ref{Eq.3}).    
\section{Partial wave analysis}
\label{PartialWaves}
This appendix presents the derivation of the scattering phase shifts for plasmons for a Berry flux target. In polar coordinates, $\mathbf{r}=(r,\phi)$, the radial and azimuthal components of the current density field $\mathbf{j}(\mathbf{r},\phi)$ can be shown to be fully determined by the charge density field $\rho(r,\phi)$ as
\onecolumngrid
\twocolumngrid
\begin{equation}
\begin{aligned}
& j_r(\mathbf{r}, \omega)=-i \frac{u^2}{\omega} \frac{\partial \rho}{\partial r}+\frac{2 u^2}{\omega_*}\Theta(r-r_{\mathrm{F}}) \frac{1}{r} \frac{\partial \rho}{\partial \phi}, \\
& j_\phi(\mathbf{r}, \omega)=-i \frac{u^2}{\omega} \frac{1}{r} \frac{\partial \rho}{\partial \phi}-\frac{2 u^2}{\omega_*}\Theta(r-r_{\mathrm{F}}) \frac{\partial \rho}{\partial r}, \label{Eq.B1}
\end{aligned}
\end{equation}
where $\omega_{*}=2n\hbar/m\mathcal{F}$ is the characteristic frequency introduced in the main text. Furthermore, due to the azimuthal symmetry of the considered problem, the charge density field obeys the closed form Bessel equation
\begin{equation}
q^2 \rho(\mathbf{r}, \omega)+\left(\partial_r^2+\frac{1}{r} \partial_r+\frac{1}{r^2} \partial_\phi^2\right) \rho(\mathbf{r}, \omega)=0,\label{Eq.B2}
\end{equation}
where $q=\omega/u$ is the wave vector of the incident plasmon. Eq.~(\ref{Eq.B2}) admits solutions in two regions, inside and outside the Berry flux target. The general form of these are written as $\rho \sim J_{\ell}(qr)$ and $\rho \sim A_{\ell}J_{\ell}(qr)+B_{\ell}Y_{\ell}(qr)$ respectively where $J_{\ell}$ and $Y_{\ell}$ are Bessel functions of the first and second kind. The coefficients $A_{\ell}$ and $B_{\ell}$ are determined by charge and current field continuity at the interface of the Berry flux target. That is, 
\begin{equation}
\begin{aligned}
& j_r^{-}(r=r_{\mathrm{F}})=j_r^{+}(r=r_{\mathrm{F}}), \\
& \rho^{-}(r=r_{\mathrm{F}})=\rho^{+}(r=r_{\mathrm{F}}). \label{Eq.B3}
\end{aligned}
\end{equation}
These boundary conditions uniquely determine the coefficients, which take the form 
\begin{equation}
\begin{gathered}
A_\ell=\frac{\pi q \ell u J_\ell(q r_{\mathrm{F}}) Y_\ell(q r_{\mathrm{F}})}{\omega_*}+1, \\
B_\ell=-\frac{\pi q \ell u J_\ell^2(q r_{\mathrm{F}})}{\omega_*}, \label{Eq.B4}
\end{gathered}
\end{equation}
which in turn determines the scattering phase shifts as $\delta_{\ell}(qr_{\mathrm{F}})=\tan(B_{\ell}/A_{\ell})$.

\section{Chiral graphene multilayers}
\label{GrapheneMultilayers}
Another promising platform for Floquet engineering involves chiral electrons in a few-layer graphene (e.g., Bernal-stacked bilayer graphene or rhombohedral trilayer graphene). The low-energy electronic states in these materials are concentrated in the vicinity of two inequivalent valleys ($\tau=\pm1$) and can be described by the following Hamiltonian:   
\begin{equation}
H(\tau,n)=
\begin{pmatrix}
    \Delta & u^\nu(\tau p_x-ip_y)^\nu\\
    u^\nu(\tau p_x+ip_y)^\nu&-\Delta,
\end{pmatrix}
\end{equation}
where $\nu=1,2,3,...$ is the number of layers in the few-layer graphene, and $u$ determines the slope of the electronic dispersion relation. 

Interestingly, the upper threshold frequency $\omega_*$ is universal and does not depend on the layer index $\nu$ or the slope parameter $u$. Really, if we combine the Berry flux $\mathcal{F}=g\nu \Delta/4\pi \epsilon_\mathrm{F}$, the collective mass is $m=\epsilon_\mathrm{F}/\nu u^{2\nu} p_\mathrm{F}^{2\nu-2}$ and electron concentration $n=g p_\mathrm{F}^2/4\pi\hbar^2 $, the resulting expression for the threshold frequency $\hbar\omega_*=2n \hbar / m \mathcal{F}=2 (\epsilon^2_\mathrm{F}-\Delta^2)/\Delta$ matches the one derived for the Dirac electrons and presented in the main text of the paper.

\section{Optical conductivity from the hydrodynamic framework and the Kubo formula}
\label{HydroKubo}
This Appendix compares expressions for the optical conductivity tensor calculated using the collisionless hydrodynamic framework and the linear response Kubo theory. For the massive Dirac model, predictions of the hydrodynamic framework for transverse $\sigma_{xy}^{\mathrm{H}}=e^{2}\mathcal{F}/\hbar$ and longitudinal $\sigma_{xx}^{\mathrm{H}}=n e^{2}/i m \omega$ conductivities (per species) can be presented as follows 
\begin{equation}
\begin{aligned}
\frac{\sigma_{xy}^{\mathrm{H}}}{\sigma_0}=\frac{\Delta}{2 \epsilon_{F}}, \quad \quad
\frac{\sigma_{xx}^{\mathrm{H}}}{\sigma_0}=\frac{\epsilon_\mathrm{F}^2-\Delta^2}{2i\hbar\omega \epsilon_\mathrm{F}}.
\end{aligned}
\end{equation}

\begin{figure}[t]
    \centering
    \includegraphics[clip,trim={0cm 0cm 0cm 2cm}, width=\columnwidth]{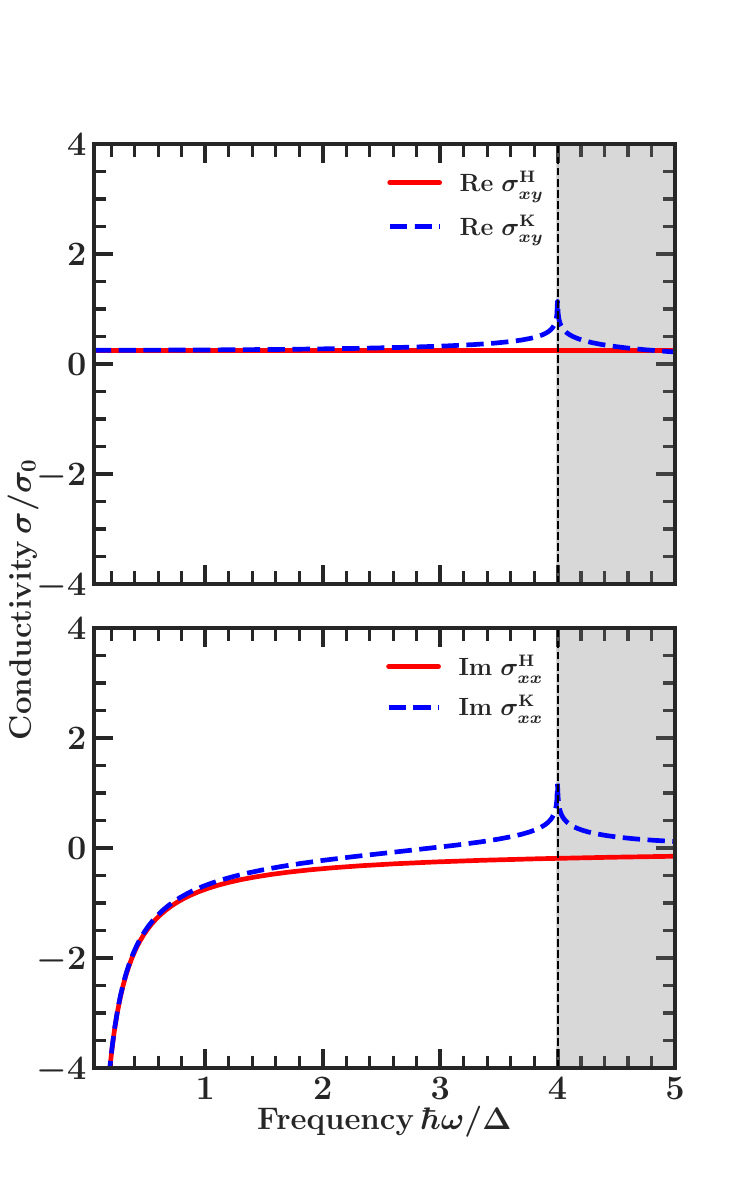}
    \caption{Transverse (top) and longitudinal (bottom) optical conductivities calculated using the collisionless hydrodynamic framework (red) and the linear response Kubo theory (blue dashed). They are calculated for alculated for $\epsilon_{\mathrm{F}}=2\Delta$ and presented in units of the conductance quanta $\sigma_{0}=e^{2}/2\pi\hbar$. The predictions of hydrodynamic theory are not only accurate over a wide frequency range, $\hbar\omega \lesssim \epsilon_{\mathrm{F}}$, but also remain reasonable up to higher frequencies, approximately $\sim 3\epsilon_{\mathrm{F}} / 2$. The hydrodynamic theory does not describe the behavior above the interband threshold, $2\epsilon_{\mathrm{F}}$, where plasmons become overdamped there and are excluded from consideration.}
    \label{Figure AP.F1}
\end{figure}

Here, $\sigma_0= e^{2}/2\pi\hbar$ is the conductivity quanta. The expressions imply $0<\Delta<\epsilon_\mathrm{F}$. According to the microscopic Kubo theory~\cite{KuboPaper}, the frequency dependence of the conductivity below the interband threshold $0<\hbar \omega < 2\epsilon_\mathrm{F}$ is given by
\begin{equation}
    \begin{aligned}
    \frac{\sigma_{xy}^{\mathrm{K}}}{\sigma_0}=&\frac{\Delta}{2 \hbar\omega}\mathrm{ln}\left( \frac{2\epsilon_\mathrm{F} + \hbar \omega}{2\epsilon_\mathrm{F} - \hbar \omega} \right),  
     \\ \frac{\sigma_{xx}^{\mathrm{K}}}{\sigma_0}=&\frac{\epsilon_{\mathrm{F}}^{2}}{2 i \hbar\omega  \epsilon_{\mathrm{F}}}-\frac{1}{8 i} \left(1+\frac{4\Delta^{2}}{\hbar^2 \omega^{2}}\right)\mathrm{ln}\left( \frac{2\epsilon_\mathrm{F} + \hbar \omega}{2\epsilon_\mathrm{F} - \hbar \omega} \right).
    \end{aligned}
\end{equation}
The frequency dependence of the optical conductivity is presented in Fig.~(\ref{Figure AP.F1}). The hydrodynamic theory not only accurately describes both longitudinal and transverse conductivities in the wide frequency range $\hbar\omega\lesssim\epsilon_{\mathrm{F}}$, but its predictions reasonably agree with the microscopic theory up to even higher frequencies $\sim 3\epsilon_\mathrm{F}/2$.  

It is worth noting that the longitudinal (transverse) conductivity develops a real (imaginary) component when the interband transition threshold $\hbar \omega > 2\epsilon_\mathrm{F}$ is exceeded. Although these effects are not accounted for by the hydrodynamic theory as it is ignorant of the actual electronic band structure of the material, and thus does not take into account interband transitions \cite{OpticalConductivityGraphene}.
\section{Eigenvalue problem reformulation}
\label{Heff}
This Appendix presents the reformulation of the hydrodynamic equations supplemented with the Poisson equation as an eigenvalue problem. If we go beyond the overscreened potential approximation, the scalar potential $\phi(\mathbf{r},t)$ created by charge density fluctuations $\phi(\mathbf{r},t)$ can be presented as
\begin{equation}
\phi(\mathbf{r}, t)=\int d \mathbf{r}^{\prime} V\left(\mathbf{r}-\mathbf{r}^{\prime}\right) \rho\left(\mathbf{r}^{\prime}, t\right). \label{Eq.C1}
\end{equation}
The potential $V(\mathbf{q})=2 \pi/q \kappa(\mathbf{q})$ incorporates the screening by surrounding media via in wavevector dependent dielectric function $\kappa(\mathbf{q})=\kappa / \tanh{kd}$ \cite{Fetter1}. Here $\kappa$ is the dielectric constant of the surrounding dielectric media, and $d$ is the distance to the metallic gate. After taking the Fourier transform of Eq.~(\ref{Eq.1}), it is instructive to introduce the spinor $\psi(\mathbf{q}, \omega)=\left\{j^{+}(\mathbf{q}, \omega), j^0(\mathbf{q}, \omega), j^{-}(\mathbf{q}, \omega)\right\}$ with chiral current components $j^{ \pm}(\mathbf{q}, \omega)=\frac{1}{\sqrt{2}}\left[j^x(\mathbf{q}, \omega) \pm i j^y(\mathbf{q}, \omega)\right]$ and a rescaled density component $j^0(\mathbf{q}, \omega)=\Omega(\vec{q}) \rho(\mathbf{q}, \omega) / q$. Here $\Omega(\vec{q})=\sqrt{2 \pi n e^2 q \tanh (q d) / m \kappa}$ is the dispersion relation of plasmons. In the main part of the paper, we consider only relatively small wavevectors $q d \lesssim 1$, where the dispersion relation is well approximated by the linear dispersion $\Omega(\vec{q})=u q$ with velocity $u=\sqrt{2 \pi n e^2 d / m \kappa}$. The resulting equations can be presented as an eigenvalue problem $\omega \psi(\mathbf{k}, \omega)=\hat{H}(\mathbf{k}) \psi(\mathbf{k}, \omega)$. The non-Hermitian $3\times3$ matrix plays the role of the effective Hamiltonian and is given by  
\begin{equation}
\label{Hamiltonian}
    \hat{H} =
    \begin{pmatrix}
    \Omega_{\mathcal{F}}(\vec{q}) & \frac{\Omega(\vec{q})e^{i\phi_{\mathbf{q}}}}{\sqrt{2}} & \Omega_{\mathcal{F}}(\vec{q})e^{2i \phi_{\textbf{q}}} \\
    \frac{\Omega(\vec{q})e^{-i\phi_{\mathbf{q}}}}{\sqrt{2}} & 0 & \frac{\Omega(\vec{q})e^{i\phi_{\mathbf{q}}}}{\sqrt{2}} \\
    -\Omega_{\mathcal{F}}(\vec{q})e^{-2i\phi_{\textbf{q}}} & \frac{\Omega(\vec{q})e^{-i\phi_{\mathbf{q}}}}{\sqrt{2}} & -\Omega_{\mathcal{F}}(\vec{q})
    \end{pmatrix},
\end{equation}
where the extra term $\Omega_{\mathcal{F}}(\vec{q})=\pi \mathcal{F} e^2 q/\hbar \kappa(\vec{q})= \Omega^2(\vec{q})/\omega_*$ originates from the nontrivial geometry of electrons participating in oscillations and is determined by the Berry flux $\mathcal{F}$. Here, $\omega_{*} = 2n \hbar / m \mathcal{F}$ is the threshold frequency introduced in the main text. 

While the Hamiltonian $H(\vec{q})$ has a non-Hermitian structure, the interplay of inversion and the particle-hole symmetries ensures its spectrum is real. Its spectrum has three eigenvalues, a flat spurious zero-frequency mode, and two dispersive modes with frequencies $\pm\Omega(\mathbf{q})$. The latter two are connected by particle-hole symmetry~\footnote{The symmetry works as $C H(\mathbf{q}) C^{-1}=-H(-\mathbf{q})$. The explicit  expression for $C$ is given by 
$$C=
\begin{pmatrix}
0 & 0 & \mathcal{K} \\
0 & \mathcal{K} & 0 \\
\mathcal{K} & 0 & 0 \\
\end{pmatrix},
$$
where $\mathcal{K}$ is the complex conjugation operator. Only the positive frequency states correspond to physical modes supported by two-dimensional electron gas.} and are not independent. The dispersion relation of plasmons $\Omega(\mathbf{q})$ is independent of the Berry flux, which enters the Hamiltonian via $\Omega_{\mathcal{F}}$. 

Due to the non-Hermitian structure of the Hamiltonian, this reformulation does not appear to provide any immediate computational or interpretative advantages for plasmon scattering.

\vspace{3cm}

\bibliography{Chiral_Plasmons}
\end{document}